\documentclass[12pt,letterpaper,twoside]{article}
\usepackage[letterpaper,dvips,body={8in,11in},vmargin={1cm,2cm},hmargin=2cm,head=1cm]{geometry}
\usepackage[dvips,usenames]{color}
\usepackage[american]{babel}
\usepackage[T1]{fontenc}
\usepackage{ae}
\usepackage{aecompl}
\usepackage{textcomp} 
\usepackage{textfit}  
\usepackage{xspace}
\usepackage{graphicx}
\usepackage{latexsym}
\usepackage{amsmath,amsfonts,amstext,amssymb,amsbsy,amsopn,eucal}
\usepackage{yfonts}[1998/10/03]
\usepackage{dsfont}
\usepackage{wasysym}
\usepackage{wrapfig}
\usepackage[normalem]{ulem}
\usepackage{url}
\newcommand\email{\begingroup \urlstyle{tt}\Url}
\urldef{\het}\url{www.het.brown.edu}
\urldef{\danieldf}{\email}{danieldf@het.brown.edu}
\urldef{\gerry}{\email}{gerry@het.brown.edu}
\urldef{\dmitri}{\email}{petrov@het.brown.edu}
\urldef{\easther}{\email}{easther@physics.columbia.edu}
\DeclareMathOperator{\Ai}{Ai}

\newcommand{\varsqrt}[1]{\ensuremath{\sqrt{#1{\,}}}\xspace}
\newcommand{\conv}[2]{\ensuremath{{#1}\ast{#2}}\xspace}

\newcommand{\dslash}{\not{\hbox{\kern-2pt $\partial$}}}
\newcommand{\pslash}{\not{\hbox{\kern-2.3pt $p$}}}
 \newtoks\nslashfraction
 \nslashfraction={.13}
 \newcommand{\nslash}[1]{\setbox0\hbox{$ #1 $}
   \setbox0\hbox to \the\nslashfraction\wd0{\hss \box0}/\box0 }

\newcommand{\plpl}{\raise-2pt\hbox{$\raise3pt\hbox{$_+$}\hskip-6.67pt\raise0.0pt
  \hbox{$^+$}\hskip 0.01pt$}}
\newcommand{\mimi}{\raise-2pt\hbox{$\raise3pt\hbox{$_-$}\hskip-6.67pt\raise0.0pt
  \hbox{$^-$}\hskip 0.01pt$}}

\newcommand{\bo}{\raise-1mm\hbox{\Large$\Box$}}              
\newcommand{\pa}{\partial}                                       
\newcommand{\trans}[1]{{#1}^{\ensuremath{\mathsf{T}}}}           
\newcommand{\nTH}{{\raise.2ex\hbox{$\displaystyle \bigodot$}\mskip-4.7mu \llap H \;}}
\newcommand{\face}{{\raise.2ex\hbox{$\displaystyle \bigodot$}\mskip-2.2mu \llap {$\ddot
        \smile$}}}                                      

\newcommand{\ev}[1]{\left\langle #1\right\rangle}        
\newcommand{\mv}[2]{\langle#1|#2\rangle}                  
\newcommand{\ip}[2]{\left\langle #1 | #2\right\rangle}    
\newcommand{\abs}[1]{\left| #1\right|}                    

\newcommand{\leftrightarrowfill}{$\mathsurround=0pt \mathord\gets \mkern-6mu
        \cleaders\hbox{$\mkern-2mu \mathord- \mkern-2mu$}\hfill
        \mkern-6mu \mathord\to$}
\newcommand{\dvec}[1]{\vbox{\ialign{##\crcr
        \leftrightarrowfill\crcr\noalign{\kern-1pt\nointerlineskip}
        $\hfil\displaystyle{#1}\hfil$\crcr}}}           

\newcommand{\sfrac}[2]{{\vphantom1\smash{\lower.5ex\hbox{\small$#1$}}\over
        \vphantom1\smash{\raise.4ex\hbox{\small$#2$}}}} 
\newcommand{\bfrac}[2]{{\vphantom1\smash{\lower.5ex\hbox{$#1$}}\over
        \vphantom1\smash{\raise.3ex\hbox{$#2$}}}}       
\newcommand{\afrac}[2]{{\vphantom1\smash{\lower.5ex\hbox{$#1$}}\over#2}}    

\newskip\humongous \humongous=0pt plus 1000pt minus 1000pt

\newif\ifdtup

\newcommand{\x}{\ensuremath{\xi}}

\begin{document}
%
%
\setcounter{footnote}{3}
\begin{titlepage}
  \begin{flushright}
    {BROWN-HET-1360; CU-TP-1089; MIT-CTP-3389 \\ June 2003} \\
    \texttt{hep-lat/0306038}
  \end{flushright}
  \bigskip
  \begin{center}
    {\Large \textbf{A Review Of Two Novel Numerical Methods in QFT}}
    \vspace{1.5cm}\\
    {\large R. Easther${}^1$, D. D. Ferrante${}^2$, G. S. Guralnik${}^{3,}$\footnote{email:
    \gerry. On leave of absence from Brown University, Providence, RI. 02912} and
    D. Petrov${}^2$} \\
    \vspace{1cm}
    {${}^1$ Department of Physics} \\
    {Columbia University, NY --- 10027, NY.} \\
    \vspace{1cm}
    {${}^2$Department of Physics} \\
    {Brown University, Providence --- 02912, RI.} \\
    \vspace{1cm}
    {${}^3$Center for Theoretical Physics} \\
    {Massachusetts Institute of Technology,
      Cambridge --- 02139, MA.}
    \date{\today}
  \end{center}
  \bigskip
  \begin{abstract}
    We outline two alternative schemes to perform numerical calculations in quantum field
    theory. In principle, both of these approaches are better suited to study phase
    structure than conventional Monte Carlo. The first method, Source Galerkin, is based
    on a numerical analysis of the Schwinger-Dyson equations using modern computer
    techniques. The nature of this approach makes dealing with fermions relatively
    straightforward, particularly since we can work on the continuum. Its ultimate
    success in non-trivial dimensions will depend on the power of a propagator expansion
    scheme which also greatly simplifies numerical calculation of traditional perturbation
    graphs. The second method extends Monte Carlo approaches by introducing a procedure to
    deal with rapidly oscillating integrals.
  \end{abstract}
\end{titlepage}
%
%
\nocite{stcqp}
\setcounter{footnote}{0}
\setcounter{figure}{2}
\section{Introduction}\label{sec:intro}

Over the last two decades, Monte Carlo numerical methods applied to
lattice quantum field theory have allowed us to perform many important
calculations, and have verified and extended our basic understanding of
elementary particle phenomena.\footnote{This document is based on a
talk given by G. Guralnik at the ``Seventh Workshop on Quantum
Chromodynamics'', 6-10 January 2003.}  While very informative, these
calculations still do not fully reflect the potential of these methods
because it has been impossible to obtain the multiple teraflop-years
of computer time required for very accurate non-quenched fermionic
calculations.  The promise of ``definitive'' lattice QCD calculations
has helped to drive the development of supercomputers to the point
where computer clusters are beginning to sustain speeds of several
teraflops. Over the next few years, many teraflop-years will be
devoted to lattice QCD, and it is likely that the range and accuracy of
prediction of QFT using numerical calculation will be greatly
extended.

However, even with this promise of great resources, numerical field
theory calculations are far from becoming mechanical. While large
blocks of computing time on fast new machines should make it possible
to deal with the fermion determinant, the effort of doing so and
dealing with lattice artifacts guarantee that it will still be
difficult to extract reasonable numerical information about fermionic
systems.  Moreover, calculations involving non-positive definite
actions, actions with rapid oscillations, phase transitions and
symmetry breaking are generally resistant or intractable to solution
by Monte Carlo methods. Many problems, such as general scattering, are
just too computationally intensive to be accessible.  Motivated by
these issues, we are in the process of developing two supplementary
approaches to traditional numerical quantum field theory. The first is
the ``Source Galerkin Method'' (SG)\cite{coco,tsgm,all} which uses
nested approximations to the Schwinger--Dyson equations. The second is
a tuned Monte Carlo method, ``Mollified Monte Carlo''
(MMC)\cite{stcqp,mmc,all} which uses an averaging process to smooth
regions of high oscillation in path integrals. Conceptually distinct,
these methods actually have similar starting points. MMC uses
information from stationary phase points of an action, while SG
methods are easiest to implement if they are iteratively constructed
around stationary phase points.

\section{Source-Galerkin Methods}

The Monte Carlo approach suffers from at least two serious intrinsic
problems associated with fermions. The first is that, in order to
apply this technique, it is essential to formulate a field theory on a
space time lattice rather than on the continuum. Any calculation, of
necessity, involves an extrapolation to the continuum for the final
result. As a consequence of the lattice, it is necessary to introduce
extra degrees of freedoms for fermionic fields.  This produces
artifacts which are difficult to control. The second problem arises
from the fact that we can not directly perform Monte Carlo sampling on
integrals with the anti-commuting Grassman variables requisite for
fermions. Consequently, any fermionic degrees of freedom must be
explicitly removed by partial integration of the lattice path
integral. These integrations result in the very non-local ``fermion
determinant'' which requires multi-teraflop computer power to evaluate
to high accuracy.

We introduced the Source Galerkin method to deal with these two
fermionic problems and the difficulties associated with phase
structure mentioned in the introduction. The SG method is based on an
iterative set of approximations to the Schwinger-Dyson equations.  It
is defined on the continuum and has the nice property that fermions
(except for anti-commutativity) are treated in the same way as
bosons. Most of the basic concepts of the SG method can be
demonstrated by its application to single field models. In particular,
we can get some understanding of why it is possible to deal with phase
structure issues. Therefore, in the interest of simplicity, we avoid
directly addressing the details of fermionic calculations here and
confine this discussion to these models. The basic ideas of how
fermions work can easily be extrapolated from a primitive lattice
precursor of this model \cite{lawson}.

We start with the simple assumption that the Quantum
Field Theory (QFT) action is written with sources $J_i(x)$ for every
field $\phi_i(x)$ so that the vacuum functional $Z = \mv{+0}{0-}_{J_i}$
satisfies the differential equation,
\begin{equation*}
  \underbrace{ F\Bigl( \frac{\delta}{\delta J_i(x)} \Bigr)}_{\substack{\text{essentially
    the} \\ \text{field equations}}}\, \mathcal{Z} = 0 \; .
\end{equation*}
A familiar but far from straightforward example of this is $\phi^4$ scalar QFT
\begin{equation*}
  (\partial^2 + m^2)\, \phi(x) + g\, \phi^3(x) = J(x) \; 
\end{equation*}
which, in Euclidean space, gives the very non-trivial set of coupled differential equations:
\begin{equation*}
  \Biggl[(\partial^2_x + m^2)\,\frac{\delta}{\delta J(x)}  +
    g\,\Bigl(\frac{\delta}{\delta J(x)}\Bigr)^3 - J(x)\Biggr] \mathcal{Z}[J] = 0 \; .
\end{equation*}

\subsection{The Ultra-Local $\boldsymbol\phi^{\mathbf{4}}$}\label{sec:ultralocal}
\begin{wrapfigure}[13]{l}{5cm}
  \begin{center}
    \vspace*{-0.8cm}
    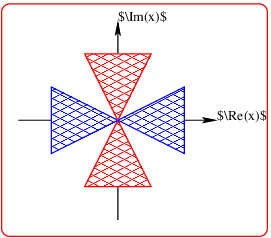
    Figure 1: Path integrals must start and end at $\infty$ in hatched regions.
  \end{center}
\end{wrapfigure}
To further simplify,  we examine the above quartic field theory in
zero dimensions. We will show that even this case has a rich basic
structure which is not directly accessible by normal Monte Carlo
approaches. Our demonstration example then reduces to the linear
differential equation
\begin{equation}
  \label{eq:schwdys}
  \Biggl[m^2\, \frac{d}{d J} + g\, \Bigl( \frac{d}{d J} \Bigr)^3 - J \Biggr]
    \mathcal{Z}[J] = 0 \; .
\end{equation}

\noindent Since this has three independent solutions, three
independent parameters are required to specify the full
solution \cite{tvbcsde}. If we rewrite $Z[J]$ as a power series,
\begin{equation*}
  Z[J] = Z[0]\, \sum_{k=0}^{\infty}\, \frac{1}{k!}\, G_k\, J^k \; ;
\end{equation*}
\noindent equation \eqref{eq:schwdys} leads to the recursion relations:
\begin{equation*}
  g\, G_{n+3} + m^2\, G_{n+1} - n\, G_{n-1} = 0 \; .
\end{equation*}

These have three independent solutions that can be expressed in terms
of parabolic cylinder functions.\cite{tvbcsde} This is a useful test
system, since it is computationally non-trivial, but has solutions in
terms of well known functions, making it is easy to check the validity
of a new numerical approximation technique. Note that, with the proper
choice of parameters, solutions exist with non-vanishing odd Green's
functions.

\begin{wrapfigure}[16]{l}{5.3cm}
  \begin{center}
    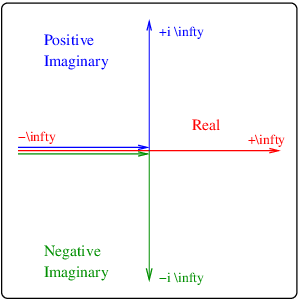
    Figure 2: Three convenient independent paths.
  \end{center}
\end{wrapfigure}

It is usual to think of the path integral formulation of QFT as being
defined by a process involving evaluation of integrals along the
entire real axis. However, this cannot yield a complete description of
the ultra local quartic theory, since it is impossible to generate
non-zero odd-order Green's functions from actions with only even
powers of fields, when path integrals are only evaluated in this
way. As a generalization of this, we see that a choice of path confined
to the real axis removes the possibility of spontaneous symmetry
breaking!  Very specifically, it is straightforward to show that the
one dimensional path integral along the whole real axis for the above
zero dimensional field theory does not produce the full solution set
outlined above.  Fortunately, it is easy to construct the path
integral formulation so that it coincides with the Schwinger-Dyson
approach \cite{tvbcsde}. For our simple model, the ``path integral''
\begin{equation*}
  Z[j] = \varint \exp\biggl\{-\frac{m^2}{2}\, x^2 - \frac{g}{4}\, x^4 + j\, x\biggr\} \,
    dx \; 
\end{equation*}

\noindent satisfies the Schwinger--Dyson equations as long as the integrand
vanishes (or the exponent goes to $-\infty$) at the ends of the
integration region. This happens if the
integration begins and ends inside the cross hatched regions
shown in  figure 1.
It follows, in agreement with our differential  analysis, that 
three independent paths can be chosen in this model,  yielding
three independent solutions. A convenient choice of paths is:
\begin{itemize}
  \item Real: $\x \in\, \mathbb{R}$;
  \item Positive Imaginary: $\x \in\, (-\infty,0]\cup[0,+i\, \infty)$;
  \item Negative Imaginary: $\x \in\, (-\infty,0]\cup[0,-i\, \infty)$.
\end{itemize}
These choices are displayed in figure 2.

Note that, because we are solving a third order linear differential
equation, the ``zero-field'',``one-field'' and ``two-field'' Green's
functions are arbitrarily determined by the choice of boundary
conditions and not by the dynamics! It is convenient to group
solutions together in 3 real combinations which can be characterized by their degree of
singularity around vanishing coupling:

\begin{itemize}
\item {Regular at $g\to 0$:} consistent with perturbation theory;
\item {$\sim g^{-1/2}$ at $g\to 0$:} ``symmetry breaking'';
\item {$\sim \exp(\mu^2/4\,g)$ at $g\to 0$:} ``instanton''.
\end{itemize}
It can easily be shown that the behavior of these solutions around
zero coupling are related to expansions around the three stationary
phase points of the path integral in the complex $\x$-plane. Indeed,
direct calculation shows that all three of these expansions form
asymptotic series corresponding to expansions of the exact solutions.

\subsection{Implementing the Source-Galerkin Approach}\label{subsec:sg}
An approximate solution to the differential equation
\begin{equation}
  D[u] = 0 \; , \label{SG_Eq}
\end{equation}
can be constructed as a linear combination of the first $N$ members of a
set of trial functions $\{\phi\}$.
\begin{align*}
  u_a &= \sum_{i=0}^{N}\, a_i\, \phi_i(x) \; .
\end{align*}
Now we can develop a procedure, which will enable us to determine
a set of coefficients $\{a\}$ so that the solution $u_a$ defined by
them is as ``close as possible'' to the exact solution for a particular
choice of $\{\phi\}$ and $N$. We start by obtaining the residue $R$ of
an approximate solution by substituting the $u_a$ into the equation
(\ref{SG_Eq}).
\begin{align*}
  R(a,x) &= D[u_a] \neq 0 \; .
\end{align*}
We define a scalar product of two functions and choose a set of weight
functions $\{w\}$. Next, a set of equations for determining $a_i$ is
constructed by requiring that the scalar product of the residue with
the weight functions vanishes:
\begin{align*}
\ip{R}{w_i} = 0, \; \forall\, i=0,\dotsc,N.
\end{align*}
This approach is an example of ``The Method of Weighted Residues''.
Provided that the $w_i$ form a complete set of functions, it guarantees
that $\lim_{N\to\infty} u_a = u$ in the mean.
If the specific choice of $w_i = \phi_i$ is made, then this approach is referred to as ``The Galerkin
Method''.

There are several points that need to be taken in to account when
using this method to solve field theories. As stated above, the
Galerkin technique produces approximate solutions which converge to
the exact solution as the number of terms goes to infinity. In
practice, the resulting series has to be truncated. Our experience to
date suggests that for nontrivial theories, calculations for $N>8$
will be impossible to perform. Consequently, when we pick functions to
describe our problem, it is essential to make a reasonable guess
consistent with all the expected properties and symmetries in order to
achieve rapid convergence, even with few terms. In practice the
apparent single pole dominance of many important qualities in physical
theories, which is so important to Monte Carlo success, seems to serve
us equally well in the Source Galerkin approach. Finally, picking a
definition of an inner product which is efficient to use with the
chosen expansion functions greatly facilitates rapidly accurate
computations.

\subsection{The Actual Numerical Solution}\label{subsec:ns}

In order to test the validity of the SG calculational technique, we applied it to the
ultra-local $\phi^4$. We solved this model using two different sets of trial functions:
powers of $J$ and a set of Hermite polynomials in $J$. The approximate solutions
constructed from these trial functions are of the form:
\begin{equation*}
  Z_a[J] = \sum_{k=0}^{N}\, \frac{1}{k!}\, a_k\, J^k \; ,\;\;\;\;
  Z_a[J] = \sum_{k=0}^{N}\, a_k\, H_n(J) \; .
\end{equation*}
The solution in terms of truncated power series is easy to deal with
and maps directly to the theoretical solution obtained in section
\ref{sec:ultralocal}. However, these are not orthogonal functions and
simple generalizations for more complicated problems do have some difficulties.
On the other hand, Hermite polynomials are orthogonal under proper choice of
inner product. This property greatly simplifies the process of
determination of the coefficients $a_i$ and ensures high numerical
stability.

The residual equations for the truncated power series were obtained using one of the inner
products presented below:
\begin{enumerate}
\item Canonical:
  \begin{equation*}
    \ip{f}{g} = \int\limits_{-\epsilon}^{\epsilon} f(J)\, g(J)\, dJ \; ;
  \end{equation*}
\item Exponentially weighted:
  \begin{equation*}
    \ip{f}{g} = \int\limits_{-\infty}^{\infty} f(J)\, g(J)\, e^{-J^2/\epsilon^2}\,dJ \; ;
  \end{equation*}
\end{enumerate}
Note that both of these inner products are chosen in such a way that
they emphasize a small region close to zero. This is done because the
generating function and its derivatives will be used to compute the
propagators which are evaluated with the sources set to zero. Epsilon
is a parameter, which allows us to tune the method to test it's
stability and achieve better numerical accuracy. It becomes highly important in
more interesting models.

\begin{figure}[Hht]
  \centering
  \scalebox{.40}{\rotatebox{-90}{\includegraphics{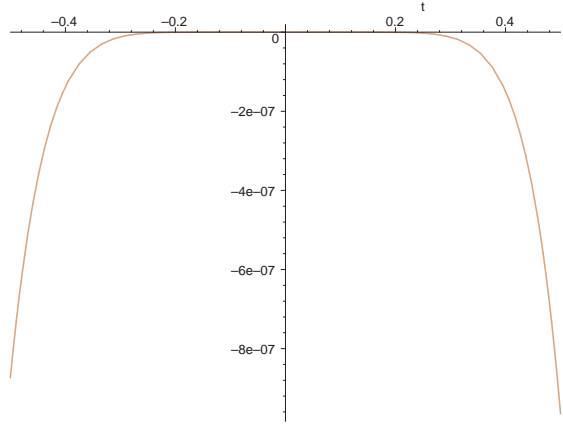}}}
  \caption{Absolute error in determination of $Z$ using a truncated power series.}
  \label{fig:power}
\end{figure}

Figure \ref{fig:power} shows dependence of absolute error in the determination of the generating
functional relative to the source $j$. It is clear that the accuracy of one part in $10^7$
is achieved and exceeded as the value of the source approaches zero.

When using Hermite polynomials as a set of trial functions, the most sensible definition of inner product
is the one under which the Hermite polynomials are orthogonal to each other.
\begin{equation*}
  \ip{f}{g} = \int\limits_{-\infty}^{\infty} f(J)\, g(J)\, e^{-J^2/\epsilon^2}\,dJ \; .
\end{equation*}

\begin{figure}[Hht]
  \centering
  \scalebox{.35}{\rotatebox{-90}{\includegraphics{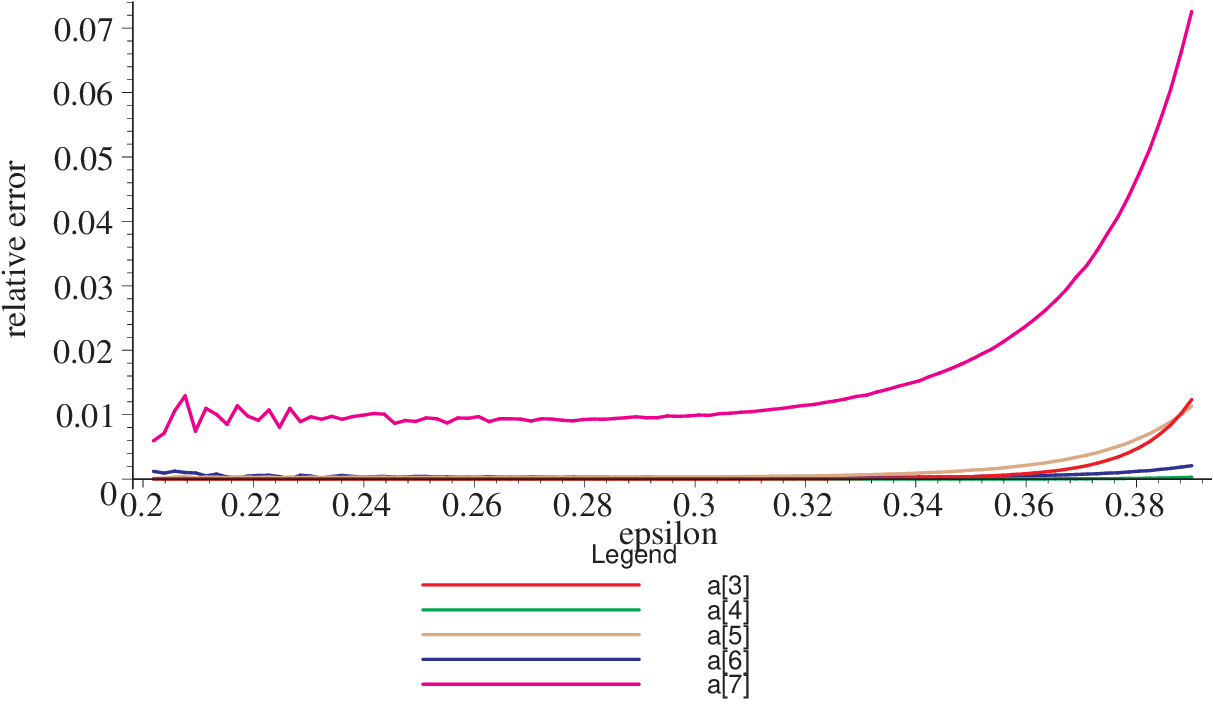}}}
  \hspace{0.4cm}
  \scalebox{.35}{\rotatebox{-90}{\includegraphics{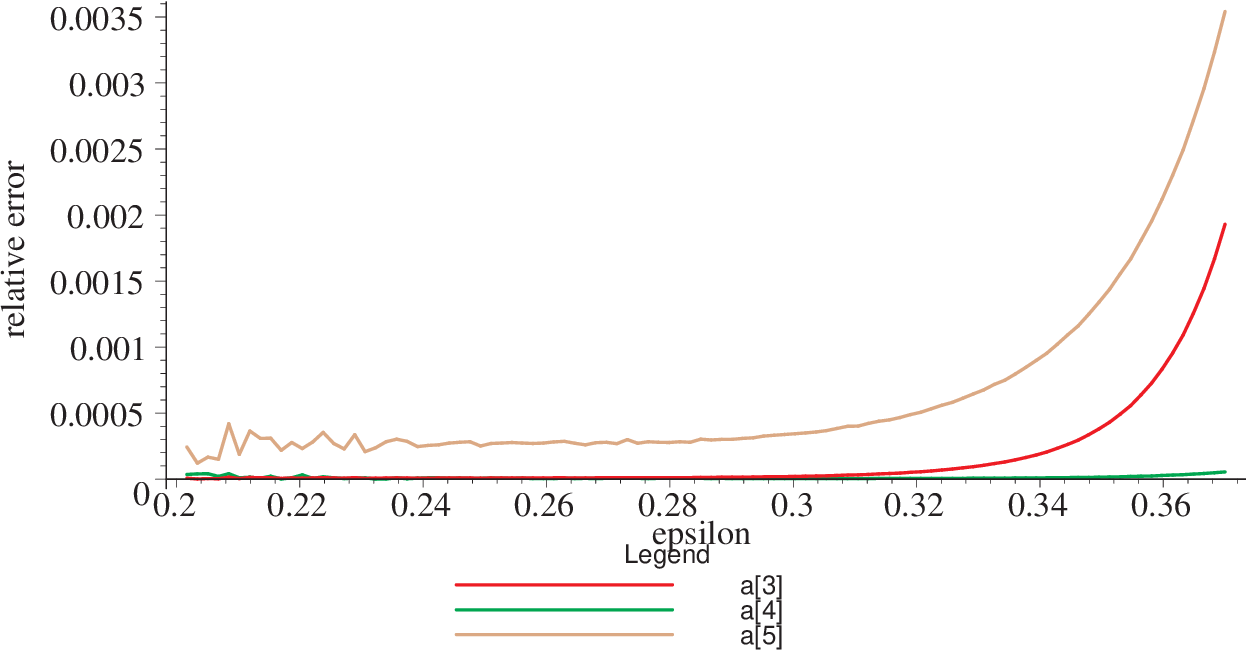}}}
  \caption{Relative error of Hermite polynomial expansions.}
  \label{fig:Hermite}
\end{figure}

The total accuracy of the computation given by this choice of trial
functions is comparable to the accuracy of the solution in terms of
the power series and is limited only by the numerical precision of the
calculation.  Order by order the relative error in determining
coefficients $a_i$ with respect to epsilon is shown on figure
\ref{fig:Hermite}. Both graphs show the same solution at different
scales.  Clearly the precision of the computation of $a_i$ decreases
at higher orders. From the graph on the left it is evident that $a_7$
is determined with an error of $1$ per cent while the graph on the right
shows that $a_5$ can be calculated more accurately than $5$ parts in
$10^4$.  This trend does not present a problem since the absolute
value of the contributions decreases at high orders thus compensating
for the loss of accuracy.

Note that the solution is constant in a significant range of values of the parameter.
At small values of epsilon this range is limited by numerical precision of the hardware used for
the computation. From the graphs it is clear that the calculation becomes numerically unstable when
epsilon is set to any value smaller than $0.24$.
For large values of the parameter, the range of stability is limited by the number of terms
in the approximate solution. If epsilon is set too high, the generating functional can not be
accurately approximated at every point in the range $(-\epsilon ... +\epsilon)$ by the linear
combination of some number (seven in this example) of Hermite polynomials.

The stability property described above is important in computations
where a theoretical solution is not available. Then we tune the parameters of a problem to
find a stable region which should indicate that we have found a good solution.

Now, we return to higher dimensions. In order to solve,
\begin{equation*}
  F\biggl(\frac{\delta}{\delta J}\biggr) Z[J]= 0\; ,
\end{equation*}
through numerically approximation, we start with the exact answer in the form:
\begin{equation*}
  Z[j] = \exp\biggl\{a + J(x)\, G(x) + \frac{J\, G^2\, J}{2!} + J\, J\, J\, G^3 + J\, J\,
    G^4\, J\, J + \dotsb\biggr\} \; .
\end{equation*}
Here, co-ordinates are only displayed in the linear term and
space-time integrations are not shown. Initially, the
$G^i(x_1,\dotsc,x_n)$ can, be anything consistent with the symmetries
of the theory. The fact that the zero dimensional model did not have
integrations is why it was so simple.

Using the Poincare invariance to produce spectral forms, we conjecture
that it should be possible to span the solution space (or a very large
part of it) with products of arbitrary two field propagators
\begin{equation*}
  G_p(x-y) = \int\frac{e^{i\, k\, (x-y)}\, a_p(K^2)}{K^2 - k^2}\, d^4k\, d^2K \; .
\end{equation*}
Thus, we represent $G^n$ by all possible graphs with $n$ external
lines, all numbers of lines intersecting at any vertex and $a_p(K^2)$
different for each propagator. Important! This is NOT PERTURBATION
THEORY. The masses and weights in each of these propagators are
arbitrary at this stage. If our conjecture is correct, we have merely
exploited the restrictions of relativity and translation invariance in
writing this representation.  The structure we have written is still
extremely complex.

The above expansion inserted into the field equations yields conditions of constraint on
$a_p(K^2)$. In general, they are too complicated to solve directly. After all, field theory
has not be exactly solved! To proceed, we simplify the problem as follows,
\begin{itemize}
\item Truncate the expansion in $J$;
\item Limit the number of masses in each propagator; 
\item Limit the number of graphs considered for each $J$.
\end{itemize}
We do this in an organized, systematic way, so that after a first
guess more terms can be included to iterate on the initial
answer. Roughly, our procedure works as follows: Begin by guessing an
initial solution $Z_{\text{approx}}[J]$ consistent with space-time,
spin and other internal symmetries. Since we have not guessed an exact
solution:
\begin{equation*}
  F\biggl(\frac{\delta}{\delta J}\biggr)\, \mathcal{Z}_{\text{approx}}[J] \neq 0 \; .
\end{equation*}
The idea of Source Galerkin is to require that,
\begin{equation*}
  \int\, dJ_1\dotsb dJ_n\, F_i(J_1,\dotsc,J_n)\, F\biggl(\frac{\delta}{\delta
    J}\biggr)\, \mathcal{Z}_{\text{approx}}[J] = 0 \; ,
\end{equation*}
so that $\mathcal{Z}_{\text{approx}}[J]$ satisfies the field equations
``on the average''. The number of (in principle arbitrary
functions) $F_i$ is picked so that all the undetermined weights in
$\mathcal{Z}_{\text{approx}}[J]$ are determined.

In general, the equations to be solved are non-linear. Consequently,
solutions must be determined in a very careful and systematic manner
and parameters must be tuned. Theorems for simpler Galerkin-like
approaches promise convergence and we assume this is true for this
approach. We have tried this method in many simple models, with
amazing accuracy when a check is available. In general, it must be
confirmed that the results are stable and convergent.  Higher order
iterations, while simple in principle, have caused us computational
difficulties in practice. While we believe that these problems can be
resolved, we have not yet fully demonstrated that our method is
practicle in more interesting theories.

\subsection{The Nonlinear Sigma Model}\label{subsec:nlsm}

Here we give a simple but interesting example of a lowest order SG
solution.  Calculation to the accuracy obtained in this case is moderately
difficult to replicate in Monte Carlo.  We have also examined leading order
corrections which, as stated, are essential for the proof of validity of the
method, and which we will discuss elsewhere.  The theory is given by
\begin{equation*}
  \mathcal{L} = \frac{1}{2}\, \Biggl\{\pa_{\mu}\,\phi^a(x)\, \pa^{\mu}\, \phi_a(x)
    + \chi(x)\biggl(\phi^a(x)\, \phi_a(x) - \frac{1}{g^2}\biggr)\Biggr\} \; .
\end{equation*}
After the introduction of sources $J^a(x)$ and $S(x)$ for canonical
and auxiliary fields, one has
\begin{equation*}
  \mathcal{L} = \frac{1}{2}\, \Biggl\{\pa_{\mu}\,\phi^a(x)\, \pa^{\mu}\, \phi_a(x) -
    J^a(x)\, \phi_a(x)
    + \chi(x)\biggl(\phi^a(x)\, \phi_a(x) - \frac{1}{g^2}\biggr) - \chi(x)\, S(x)\Biggr\}
    \; .
\end{equation*}
The equations of motion that follow are:
\begin{align*}
  \Box\, \frac{\delta Z[J,S]}{\delta J^a(x)} - \frac{\delta^2 Z[J,S]}{\delta S(x)\, \delta
    J^a(x)} -J^a(x)\, Z[J,S] &= 0 \; ;\\
  \frac{\delta^2 Z[J,S]}{\delta J^a(x)\, \delta J^a(x)} - \frac{1}{g^2}\, Z[J,S] - 2\,
    S(x)\, Z[J,S] &= 0 \; .
\end{align*}

We guess the leading approximation to be
\begin{equation*}
  Z_a[J,S] = Z_0\, \exp\Biggl\{\frac{1}{2}\, \int J^a(x)\, G^{a\,b}(x-y)\, J^b(y)\, dx\,dy
    + \int \chi_0\, S(x)\, dx \Biggr\} \; .
\end{equation*}
The residues are computed by substituting this expression into the Schwinger Dyson
equations,
\begin{align*}
  R_1 &= \Box \int G^{a\, b}(x - \xi)\, J^b(\xi)\, d\xi - J^a(x) - \chi_0\int G^{a\, b}(x
    - \xi)\, J^b(\xi)\, d\xi \;  \\
  R_2 &= G^{a\, a}(x - x) + \int G^{a\, b}(x - \xi)\, J^b(\xi)\, d\xi\, \int G^{a\, b}(x -
    \eta)\, J^b(\eta)\, d\eta - \frac{1}{g^2} - 2\, S(x) \; .
\end{align*}

The Source Galerkin equations can be obtained by requiring that the projections of the
residues on the weight functions vanish:

\begin{align*}
  \ip{1}{\ip{J^b(y)}{R_1(x)}_J}_S &= 0 \; ; \\
  \ip{1}{\ip{1}{R_2}_J}_S &= 0 \; ;
  \intertext{which yield,}
  \Box\, G^{a\, b}(x-y) - \delta^{a\, b}(x-y) - \chi_0\, G^{a\, b}(x-y) &= 0 \; ; \\
  G^{a\, a}(0) + \epsilon^2\, \int G^{a\, b}(\xi-\eta)\, G^{b\, a}(\eta-\xi) -
    \frac{1}{g^2} &= 0 \; ;
\end{align*}
where, the following definition was used for the inner product
\begin{equation*}
  \ip{A}{B}_J = \frac{1}{C_{\epsilon}}\, \int A[J(x)]\, B[J(x)]\,
    e^{-J^2(x)/\epsilon^2}\, [dJ] \; .
\end{equation*}
It is straightforward to show that the results obtained here give the
same two field Green's function as the leading order large N
expansion.

As indicated, the problems with our method occur at
higher orders in the expansion.  Since our approach can be represented
by arbitrary combinations of Feynman graphs whose masses and
coefficients are set by solving Galerkin equations, fast numerical
evaluation of these graphs is essential for Source Galerkin to be
useful. It is for this reason that we developed a new method to
numerically study Feynman graphs.

\subsection{Feynman Diagrams}

Motivated by our work on the Source Galerkin technique, we have
carefully looked at the numerical calculation of Feynman
graphs. Historically complicated graphs are evaluated with Monte
Carlo methods which take a relatively long time to achieve accurate
results. We have have been able to devise a new method for numerically
evaluating graphs which is very accurate and very fast, and has
considerable promise for complex traditional perturbation calculations
(such as the eighth order magnetic moment). The availability of such a
method is crucial to fully implementing our Galerkin approach.

We use an approximation to the propagator that reduces graph
evaluation to a rapidly convergent multi-dimensional sum.\cite{fefd}
We construct a propagator representation as follows: The
cutoff two field propagator, $G_{\Lambda}$
\begin{equation*}
  G_{\Lambda}(x) = \int\frac{d^4p}{(2\, \pi)^4}\; \frac{e^{i\, p\, x}}{p^2 +
    m^2}\, e^{-p^2/\Lambda^2} \;
\end{equation*} 
is written using the Sinc function,
\begin{equation*}
  S_k (h,x) \equiv \frac{\sin(\pi\, (x - k\, h)/h)}{\pi\, (x - k\, h)/h}\; ,\quad
    k\in\mathbb{Z} \;
\end{equation*}
as
\begin{align*}
  G_{\Lambda,h}(x) &= \frac{m^2\, h}{(4\, \pi)^2}\, \sum_{k=-\infty}^{\infty}p(k)\, \exp\biggl[-\frac{m^2\,
    x^2}{4\, C(k)}\biggr] \\
  & \\
  C(k) &= e^{k\, h} + \frac{m^2}{\Lambda^2} \;,\;
  p(k) = \frac{e^{k\,h - e^{k\, h}}}{C^2(k)} \; .
\end{align*}
The accuracy of the approximation is tuned by the value of $h$, and
typically, the propagator can be approximated to very high accuracy (1
part in $10^{16}$) with fewer than 100 terms in the sum. Inserting
this representation of the propagator into the graphical topologies we
wish to calculate ensures that all the space-time or momentum
integrals are Gaussians and can be performed with ease. The resulting
multi-dimensional sum can then be quickly and accurately computed
numerically.  We have calculated 4th non-trivial order (3 loops) in
scalar field theory\cite{master} and successfully responded to
challenges to the accuracy and speed of the method. We have computed
the QED graphs shown in the diagram on the right\cite{em} as well as a
few higher order graphs.
\begin{wrapfigure}[9]{r}{5.5cm}
  \begin{center}
    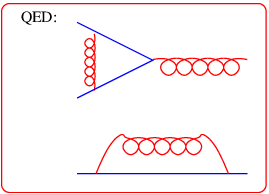
  \end{center}
\end{wrapfigure}

Technical issues with accuracy using an ``auto'' renormalization
scheme have slowed the development of a fully automated implementation
of this method to higher order QED graphs. We believe these
difficulties are soluble.  Moreover, the power of this graphical
calculation method gives us reason to believe that we can iterate the
SG method to moderate order with relatively small amounts of computer
time. If this is true, a much broader range of QFT will be opened to
numerical solution with current computer power. 

Finally, we note that there is a long history of trying to solve QFT
by making approximations to the Schwinger Dyson equations. Our
approach is unique because of our way of picking best fits and our
systematic procedure based on the powerful way of evaluating arbitrary
graphs using the Sync expansion techniques described above. The current power of
computers has made it possible to revisit these ideas and institute
algorithms which previously would have been useless.

\section{Mollified Monte Carlo}\label{sec:mmc}

The success of Monte Carlo methods is very dependent on the form that
the ``action'' takes. If some terms are not positive (sign problem) or
imaginary, or if the action has regions of very rapid oscillation,
conventional Monte Carlo approaches will mostly fail. The failed
actions are amongst the more interesting ones that we would like to
evaluate. As a concrete example, consider symmetry breaking such as
was discussed in the SG sections. We saw that to get non-vanishing odd
Green's functions, it was necessary to extend the path integral onto
the imaginary axis or, equivalently, to examine actions with negative
bare mass terms and half line integrations. In general we would not
expect Monte Carlo to be able to handle this. Further it would be very
nice if we could, at least in some formal manner, examine actions in
Minkowski space, which means we must deal with (carefully defined)
complex integrals. Again, we can not do this conventionally.  The
study of phase structure means looking at actions in regions of high
oscillation. Using normal approaches,this is asking for trouble. 

Doll and his collaborators \cite{stcqp,mmc} have discovered a way of
rewriting oscillatory integrals which has the potential to overcome
most of these problems and allow Monte Carlo like numerical
evaluation. We will outline some of the fairly complex details of this
approach in what follows, but it is straightforward to state the
general procedure. Basically, we will show by the use of a probability
function it is possible to rewrite (mollify) an oscillating path
integral, so regions that do not contribute (regions of non-stationary
phase) are effectively removed. Next we sample the new integral form
using importance functions which are heavily weighted around the
stationary phase regions of the mollified integral. This modified
Monte Carlo approach will converge even with integrals which
originally possessed awful oscillations. In fact, Mollified Monte
Carlo (MMC) tends to yield results near to mean field (large N)
analytic approaches.  The results will also be close to SG expansions
picked in our usual way.  We have examined this ``mollified'' approach
extensively in zero dimensional field theory and are now trying to use
it to solve real theories. While we see rapid convergence in regions
of high oscillation, this method is more complex than normal Monte
Carlo and will not save any time when used in non-oscillatory
problems. Further, it provides no new insight into unquenched
fermionic calculations - only the potential of convergence in regions
previously not accessible.

We make this more specific by showing some simple examples.  Our
starting point is a seemingly innocent identity using a probability
distribution, We define:
\begin{align}
  \label{eq:equality}
  \int\, f(x)\, dx &\mapsto \int\, \ev{f(x)}_{\epsilon}\, dx \; , \\
  \intertext{where}
  \label{eq:def}
  \ev{f(x)}_{\epsilon} &\equiv (\conv{f}{P_{\epsilon}})(x) \; , \\
  \nonumber
  &= \int\, P_{\epsilon}(x-y)\, f(y)\, dy \; ,
  \intertext{with}
  \nonumber
  \int\, P_{\epsilon}(x)\, dx &= 1 \; .
\end{align}
In this case, the smoothing function is given by $P_{\epsilon}(x)$ which is
called a \emph{``mollifier''} or an \emph{``approximate identity''}. We use one
of the properties of the mollifiers,
\begin{equation*}
  \label{eq:limit}
  \ev{f(x)}_{\epsilon} \stackrel{\epsilon \to 0}{\longrightarrow} f(x) \;
    \Rightarrow\; \int\, \ev{f(x)}_{\epsilon}\, dx \stackrel{\epsilon \to
    0}{\longrightarrow} \int\, f(x)\, dx \; .
\end{equation*}
With this simple averaging method we can build \emph{mollified}
versions of the desired partition functions and Green's functions,
tame oscillations and evaluate previously numerically
inaccessible functions.

\subsection{Simple example}

Let us show how oscillations can be tamed by mollifying the very singular function 
$\exp(i/x) = \cos(1/x) + i\, \sin(1/x)$.
\begin{align*}
  g(x) = \exp(i/x) &\; , \quad P_{\epsilon}(x) = \frac{\exp\Bigl\{-\frac{1}{2}\,
    \frac{x^2}{\epsilon^2}\Bigr\}}{\varsqrt{2\, \pi\, \epsilon^2}} \\ &\\
  \therefore\; \ev{g(x)}_{\epsilon} = \int_{-\infty}^{\infty}\,&
  \frac{\exp\bigl\{-(1/2)\, (x - y)^2/\epsilon^2\bigr\}\, \exp\{i/y\}}{\varsqrt{2\,
      \pi\, \epsilon^2}}\, dy
\end{align*}

\begin{center}
  \scalebox{0.3}{\includegraphics{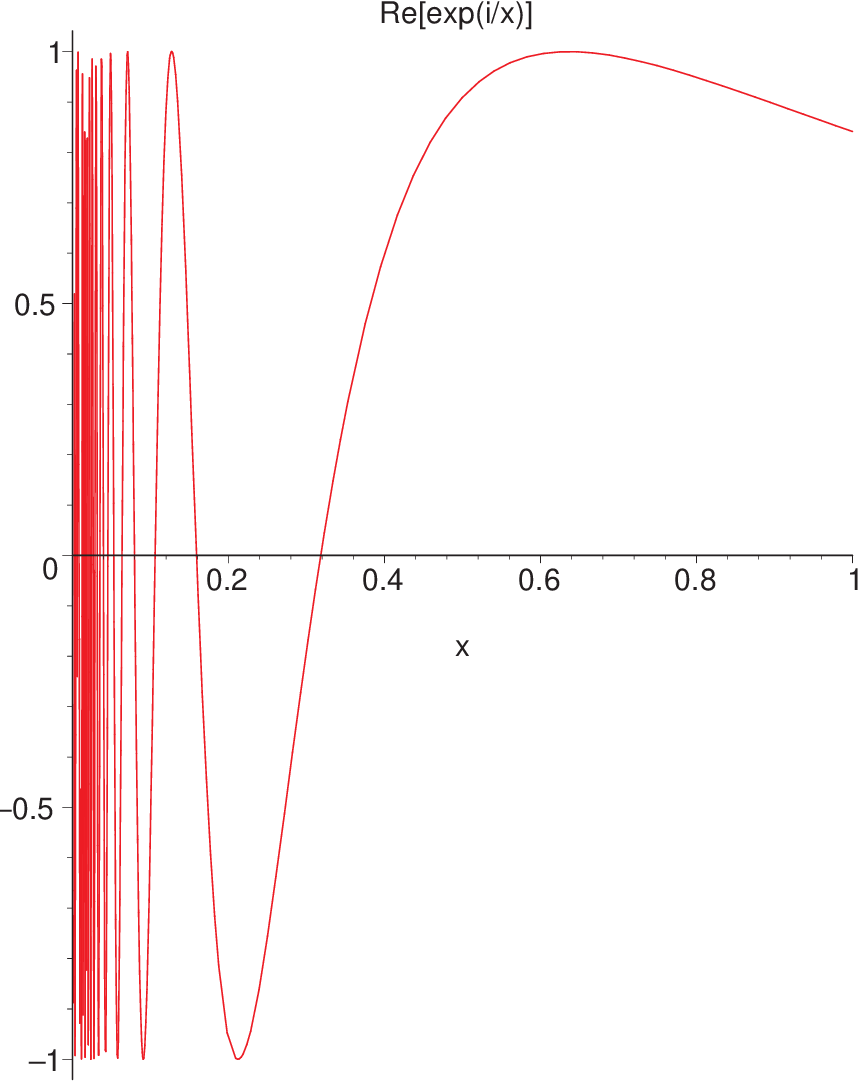}}
  \hspace{1cm}
  \scalebox{0.3}{\includegraphics{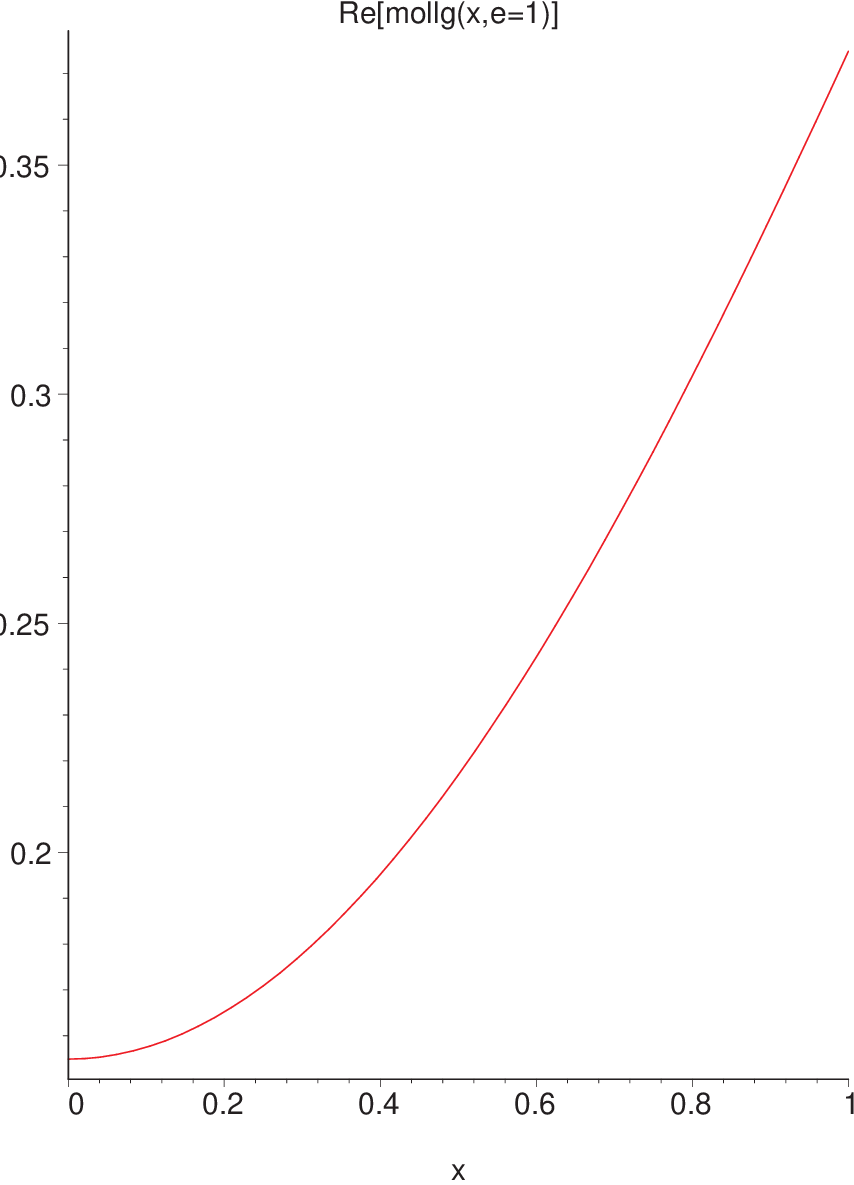}}
  \hspace{\fill}
\end{center}
An analogous behavior is seen for the imaginary part. It is quite clear that we have smoothed
an essential singularity. We will show that this can be used without loosing information
to evaluate Green's functions.

\subsection{Simple example (3D): Gaussian mollifier}

We give one more example to show that the above result was not a unique case:

\begin{align*}
  f_m(x) = \exp(i\,m\,x^2) &\; , \quad P_{\epsilon}(x) = \frac{\exp\Bigl\{-\frac{1}{2}\,
    \frac{x^2}{\epsilon^2}\Bigr\}}{\varsqrt{2\, \pi\, \epsilon^2}} \\
  & \\
  \Rightarrow \quad \ev{f_m(x)}_{\epsilon} &=
  \frac{\exp\Bigl\{\frac{i\,m\,x^2}{1 - 2\,i\,m\,\epsilon^2}\Bigr\}}{\varsqrt{1 -
      2\,i\,m\, \epsilon^2}}
\end{align*}

\begin{center}
  \hspace{\fill}
  \scalebox{0.3}{\includegraphics{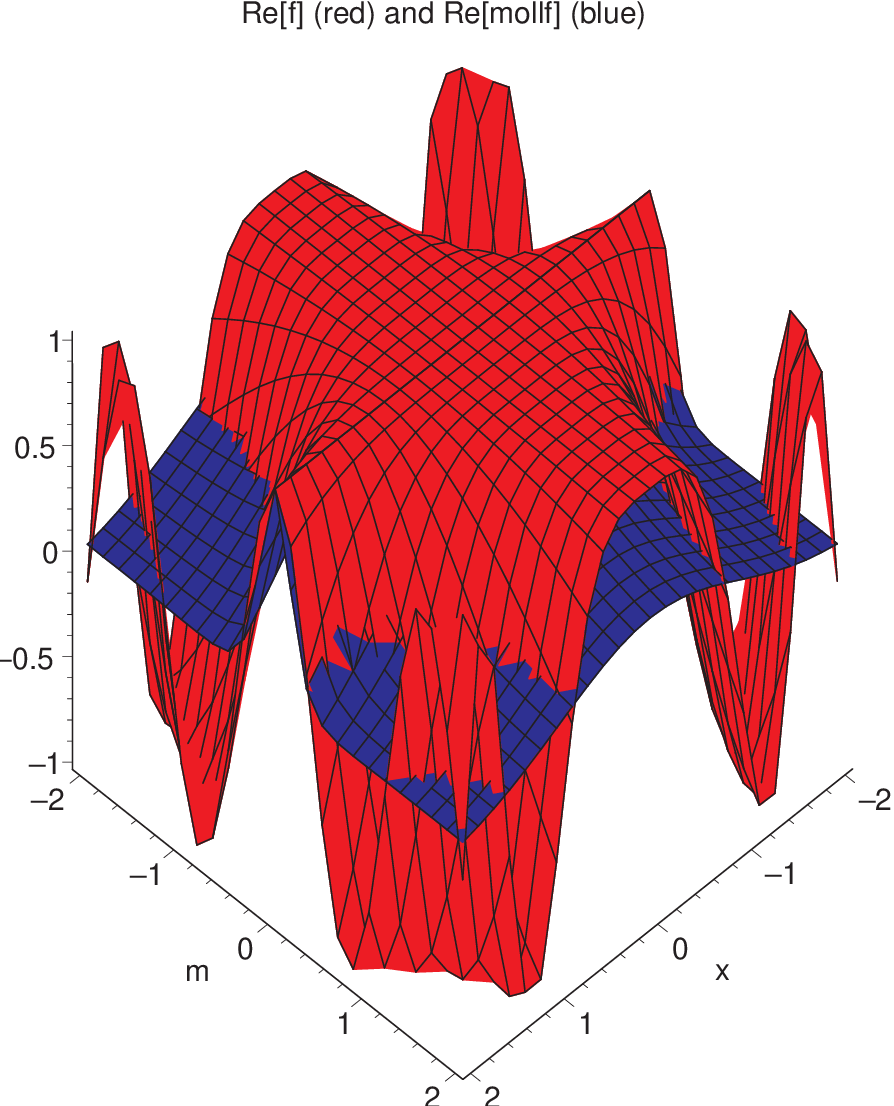}}
  \hspace{\fill}
  \scalebox{0.3}{\includegraphics{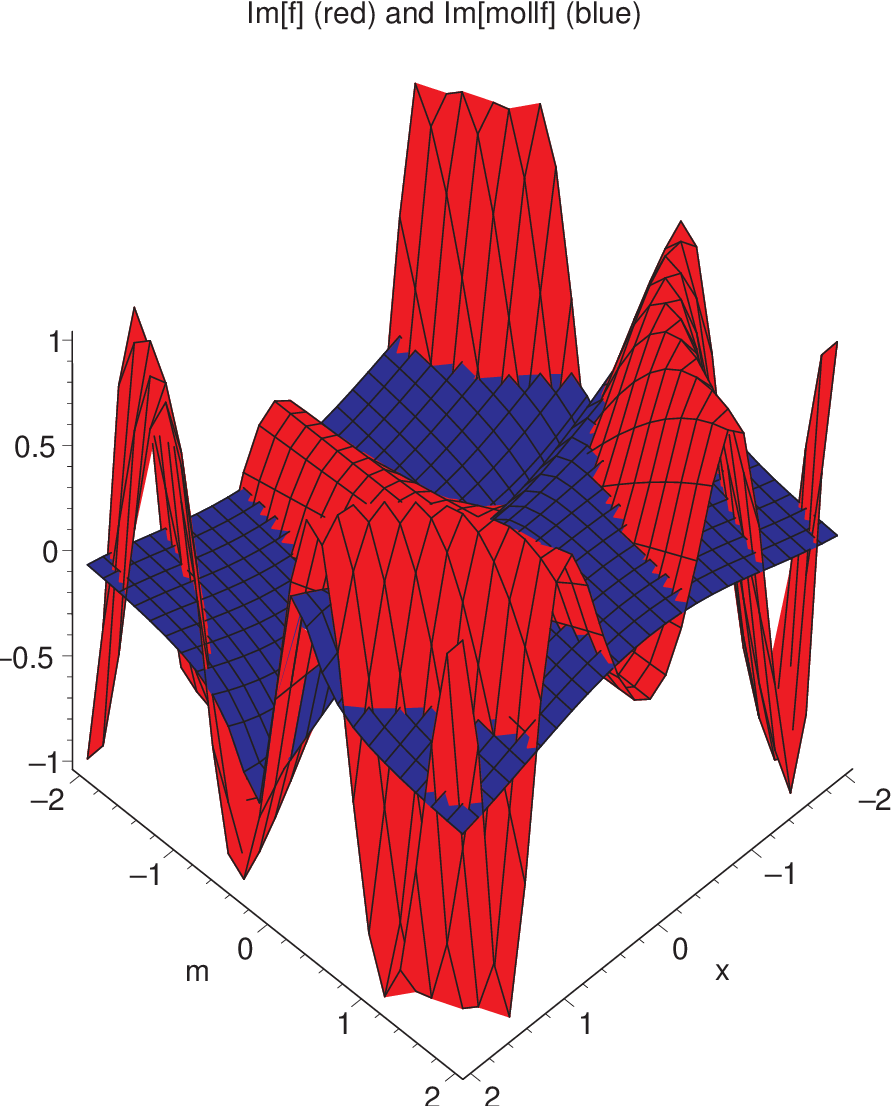}}
  \hspace{\fill}\hspace{\fill}
\end{center}
Once again, it is easy to see that the surfaces in blue/dark grey (mollified)
are less oscillatory than their counterparts (red/light grey). This is quite
useful when one starts to think about combining the mollifier
technique with the Monte Carlo one and sees that we have tamed this
highly oscillatory function for positive and negative values of mass
and for real and imaginary contributions. This makes it clear that we
should be able to handle general sign problems.

\subsection{Application of the Method}

Now we can outline the application to Monte Carlo integrals.
The method, consists of two basic steps. We start by mollifying (pre-averaging) the
integrand. The generating functional is given by,
\begin{equation*}
  \mathcal{Z}_{\epsilon}[j] = \frac{\displaystyle\int\, \ev{e^{i\, S(\mathbf{x}) -
        i\,\mathbf{j\cdot x}}}_{\epsilon}\, [dx]}{\displaystyle\int\, \ev{e^{i\,
        S(\mathbf{x})}}_{\epsilon}\, [dx]} \;\; ,
\end{equation*}
using the following [Gaussian] mollifier:
\begin{align}
  \nonumber
  P_{\epsilon}(\mathbf{y}) &= \frac{\exp\Bigl\{-\frac{1}{2}\, \trans{\mathbf{y}}\cdot
    (\boldsymbol\epsilon^2)^{-1}\cdot \mathbf{y} \Bigr\}}{\varsqrt{(2\pi)^n\, \det(\boldsymbol\epsilon^2)}} \; .\\
  \nonumber
  & \\
  \intertext{We next perform a \emph{stationary phase expansion which produces the
    structure we will use for the importance sampling functions}. We start with the basic
    integral}
  \nonumber
  \mathcal{I}_{\epsilon} &= \int \ev{f(\mathbf{x})\, e^{i\,S(\mathbf{x}) - i\,\mathbf{j\cdot x}}}_{\epsilon} \,
    [dx] \; , \\
  \nonumber & \\
  \intertext{and derive the result,}
  \label{eq:2gradapprox}
  \mathcal{I}_{\epsilon} &\approx f(\mathbf{x_0})\, e^{i\,S(\mathbf{x_0}) - i\,\mathbf{j\cdot x_0}}× \\
  \nonumber
  &× \int\, \frac{\exp\Bigl\{-\frac{1}{2}\, \trans{\mathds{B}}\cdot (\mathds{1 +
    \trans{\boldsymbol\epsilon}\cdot \mathds{S}''\cdot \boldsymbol\epsilon})^{-1}\cdot
  \mathds{B}\Bigr\}}{\varsqrt{\det(\mathds{1} + \trans{\boldsymbol\epsilon}\cdot \mathds{S}''\cdot
  \boldsymbol\epsilon)}} \, [dx] \; ,
\end{align}
where $\boldsymbol\epsilon^2$ is the covariance matrix that defines
the Gaussian distribution, $\mathds{B} = \boldsymbol{\epsilon\cdot
\nabla} S(\mathbf{x})$ and $[\mathds{S}'']_{ij} = \partial^2
S(\mathbf{x})/\partial x_i\, \partial x_j$. Note, we have suppressed
the important point that, in general, the functions under study will
have multiple stationary phase points.

The integrals of interest take the form:
\bigskip
\begin{equation}
  \label{eq:impsample}
  \mathcal{Z}_{\epsilon}[j] = \frac{\displaystyle\int\, \Biggl[\frac{\ev{e^{i\,
          S(\mathbf{x}) - i\, \mathbf{j\cdot x}}}_{\epsilon}}{W(\mathbf{x})}\Biggr]\, [d
    W(\mathbf{x})]}{\displaystyle\int\, \Biggl[\frac{\ev{e^{i\,
          S(\mathbf{x})}}_{\epsilon}}{W(\mathbf{x})}\Biggr]\, [d W(\mathbf{x})]} \;\; ,
\end{equation}
where $W(\mathbf{x})$ is known as the \emph{importance function} and the MC (Metropolis)
sampling is done over this new measure. A reasonable properly biased choice is given by,
the result of our saddle-point expansion. Thus:
\bigskip
\begin{equation*}
  \label{eq:impsampchoice}
  \hspace{-2cm}
  W_{\epsilon}(\mathbf{x}) = \abs{
    \frac{\exp\Bigl\{i\,S(\mathbf{x_0}) -\frac{1}{2}\, \trans{\mathds{B}}\cdot (\mathds{1 +
        \trans{\boldsymbol\epsilon}\cdot \mathds{S}''\cdot
        \boldsymbol\epsilon})^{-1}\cdot \mathds{B}\Bigr\}}{\varsqrt{\det(\mathds{1} +
        \trans{\boldsymbol\epsilon}\cdot \mathds{S}''\cdot \boldsymbol\epsilon)}} }
\end{equation*}

\subsection{Airy Function}\label{subsec:af}

The action for this model is given by: $S(x) = x^3/3 + t\, x$. This is
an interesting model because one can explicitly calculate the
partition function and compare it with the results coming from our
mollified Monte Carlo procedure. Moreover, this model has two
stationary phase points and the results we display are from the one in
the complex plane which is not accessible with normal Monte Carlo. The
quantity of interest is:
\begin{equation*}
  \mathcal{Z}[t] = \frac{\displaystyle\int_{-\infty}^{\infty}\, \exp\biggl\{i\,
    \frac{x^3}{3} + i\, t\, x\biggr\}\, dx}{\displaystyle\int_{-\infty}^{\infty}\,
    \exp\biggl\{i\, \frac{x^3}{3}\biggr\}\, dx} \equiv \frac{\Ai(t)}{\Ai(0)} \; .
\end{equation*}

The graphs below show the behavior of the partition function. In the leftmost one, the
oscillatory behavior can be clearly seen and, in the rightmost one, the same behavior is
shown (for a fixed $t=-16$, in blue/dark grey) together with the mollified version of the partition
function (in red).

\begin{center}
  \hspace{\fill}
  \scalebox{0.25}{\includegraphics{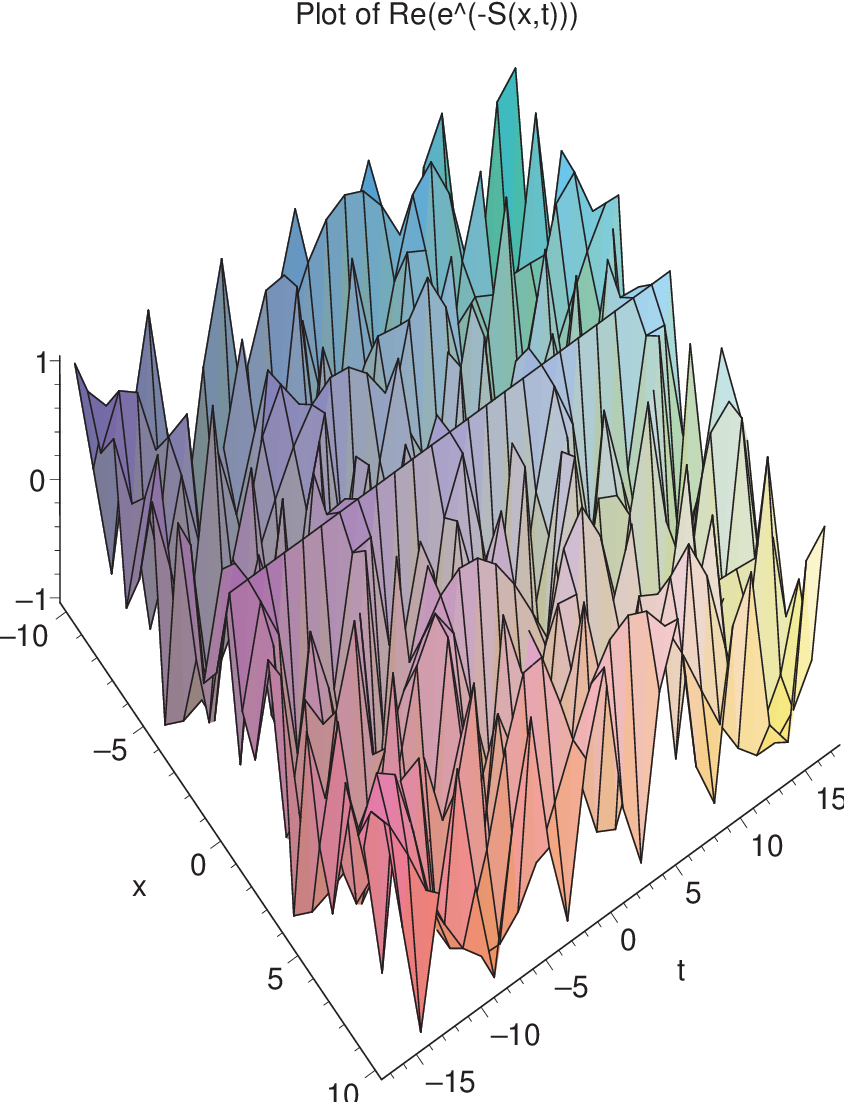}}
  \hspace{\fill}
  \scalebox{0.25}{\includegraphics{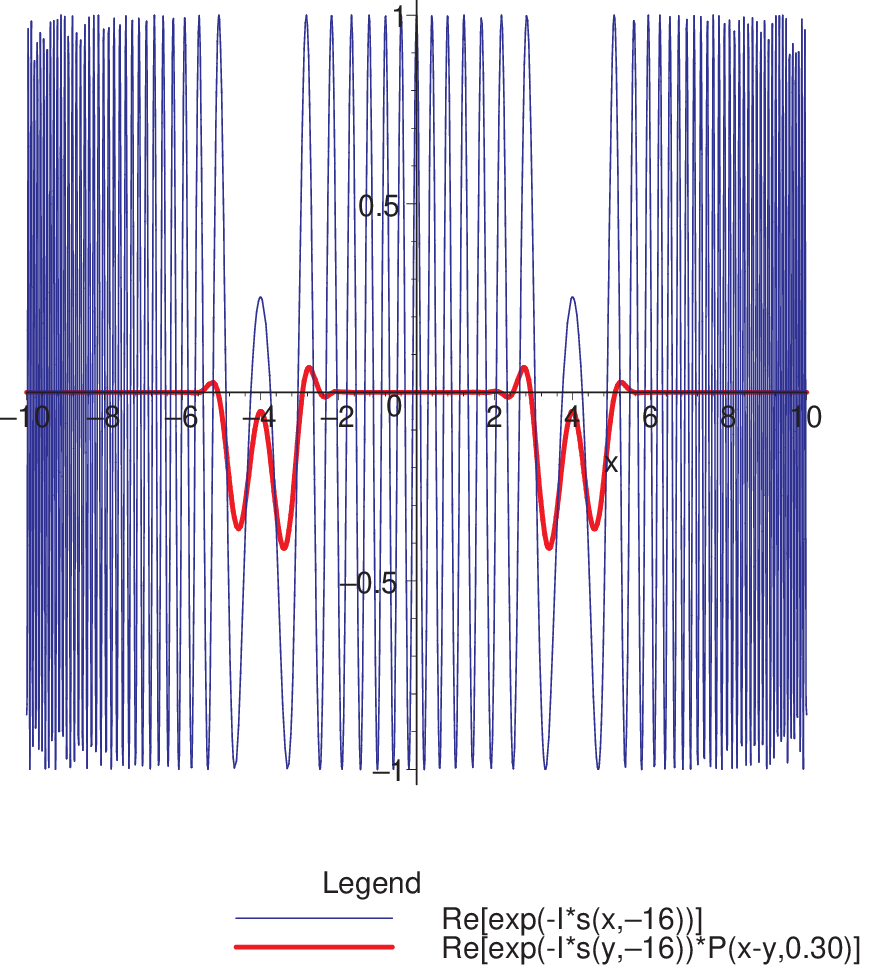}}
  \hspace{\fill}\hspace{\fill}
\end{center}

It is easy to perform the detailed calculations we outline above to get the pure numerical MMC
evaluation of this integral. The results are sensationally accurate and show that in this simple
example that we can achieve results not allowable in ordinary Monte Carlo approaches. Further, we
have done these calculations for the quartic ultra local theory and achieve similar highly accurate results
corresponding to all three solutions.

\section* {Conclusions}

We have examined two possible numerical methods that could extend the
results of normal Monte Carlo approaches. We have demonstrated that, at
least in zero dimensional field theory, these methods make sense
and have unique and useful properties. We only have glimpses of the
validity of either method in higher dimensions. However, the
development of the higher dimensional SG analysis has led to a new and
very rapid method of doing normal perturbation theory. We expect to be
able to push both of our methods considerably further in the next
year. However, full confirmation of their usefulness will
probably have to wait for work by other groups since, like all
numerical quantum field theoretic calculations, large collaborations will be
required for hard calculations.

\section*{Acknowledgements}
This talk is largely a review of previously reported research . We
would like to thank many of our colleagues who have participated in
major parts of the work described here (see references). They include Jimmie
Doll, Pinar Emirdag, Zack Guralnik, Stephen Hahn and D. Sabo.

G.G acknowledges hospitality at the Center for Theoretical
Physics-MIT. This work is supported in part by funds provided by the
US Department of Energy (DOE) under co-operative research agreements
DF-FC02-94ER40818 and DE-FG02-91ER40688-TaskD

%
%
\bigskip
%

%
\end{document}